## Title

**Designing Refractive Index Fluids of Food Dye for Light Propagation through Scattering Media**

## Authors

Muhammad Waqas Shabbir[1†], Sagor Biswas[2†], Rohit Kajla[1], Sahithi Nadella[1], Zihao Ou[1,2*]

[1]Department of Physics, The University of Texas at Dallas, Richardson, TX 75080

[2]Department of Electrical and Computer Engineering, The University of Texas at Dallas, Richardson, TX 75080

[†]These two authors contributed equally.

[*]Corresponding-author Email: Zihao.Ou@UTDallas.edu

## Abstract

Scattering and absorption are fundamental processes in optical engineering and applications. This study investigates the use of the food dye tartrazine to design refractive index fluids that enhance light propagation through scattering media. The optical properties of the solutions were carefully examined using spectrometry and ellipsometry under two extreme conditions, emphasizing the importance of a comprehensive understanding of dye solutions. Additionally, we demonstrated using dye molecules to control light propagation through scattering media. Our findings highlight the potential of food dyes as cost-effective, environmentally friendly alternatives for future advancements in biomedical imaging, optical communication, and photonic devices.

## Key words

## Introduction

Scattering and absorption of materials have been pivotal in optical engineering, with applications spanning from biomedical imaging to optical communication devices.[1,2] These fundamental processes influence how light interacts with materials, affecting the propagation of light through the optical systems.[3] Understanding and controlling these interactions are crucial for advancing technologies and the refractive index of materials is a fundamental parameter that determines their optical properties, which determines how light bends and propagates through a medium, influencing phenomena such as reflection, refraction, and scattering. By manipulating the refractive index, the optical behavior of materials can be tailored to achieve properties exceeding natural materials and such strategies have been utilized in designing dynamic optoelectronic and optofluidic devices. By manipulating the neighboring refractive index environment surrounding nanoscopic features, the scattering properties of the flat-optic platform can be drastically tuned, opening up unique possibilities in dynamic display and sensing applications.[4]

Strongly absorbing molecules can dramatically alter the optical properties of materials. Although strongly scattering, these dye molecules can increase refractive index in certain wavelengths simply by loading into the parent materials of solid or liquid form.[5,6] For example, introducing dye molecules into aqueous solutions can tune the refractive index of solutions at specific wavelength ranges,[5] and they can also be incorporated in the preparation of polymeric materials to enhance the scattering of light in selected wavelength.[6] Based on the fundamental connections between absorption and refractive index, the scattering properties of the materials can be also engineered without changing the original composition of the original material. This principle has led to the innovative solution of changing the scattering properties of tissue in living animals, where a dye molecule have been introduced into the biological tissues, such as skin, to reduce the scattering from the originally-opaque materials.[7] Such fundamental insight promises to enhance the comprehension of the underlying principles light-tissue interaction and to provide new perspective for developing innovative methods in optical clearing.[8]

Food dyes have been a unique category of strongly absorbing molecules. Food dyes offer several advantages, including cost-effectiveness, wide availability, and safety.[9-11] Unlike traditional optical materials, food dyes are relatively non-toxic and environmentally friendly, making them suitable for a broader range of applications. By carefully selecting and mixing food dyes with host materials, it is possible to design materials with precise refractive indices and absorption characteristics, that is drastically different from the original host materials.[5-7] Natural systems have also adopted similar strategies to tune the refractive index of the liquid materials to filter selected wavelength and enhance the focusing capabilities of eye at the same time utilizing dye-containing oil droplets.[12]

This research aims to explore the potential of food dye-based refractive index fluids for light propagation through scattering media. We investigate the preparation, characterization, and demonstrate how these fluids can be utilized to engineer scattering properties of the model composite systems. The unique capabilities to decouple the optical properties of the host materials from other physical properties give unique potential in optical engineering, paving the way for significant advancements in optical technologies, particularly in fields where precise control of light propagation is essential. This includes not only biomedical imaging and optical communication but also areas such as environmental monitoring, where accurate light measurement is critical, and in the development of new photonic devices, where innovative materials can lead to breakthroughs in technology.

## Materials and Methods

### *Statement of human and animal rights*

This research does not involve any human or animal subjects.

### *Dye Solution Preparation and Characterizations*

Dye solutions were prepared by mixing the dye powder with the deionized water as detailed in **Table 1**. To prepare solutions of different concentrations, each sample was made by dissolving an appropriate quantity of dye powder in deionized water from a water purification system (SOLO$^{TM}$ S, Avidity Science). Without further clarification, the dye solid is referring to tartrazine powder ($C_{16}H_9N_4Na_3O_9S_2$, CAS number: 1934-21-0, molecular weight: 534.36 g/mol, ≥85 %, MilliporeSigma) and the powder were used without further purification. To ensure accuracy in weight measurement, all weight measurement were carefully measured with a high precision balance (0.1 mg accuracy, HZK Professional Analytical Balance) and the volume of solution were measured using volumetric pipettes (Eppendorf Research plus BASIC). Two batches of samples were prepared for different purposes: (1) Low concentrations (<0.01%) of tartrazine for extraction the extinction coefficient of the molecules; (2) High concentrations (>5%) for ellipsometry measurements to quantify the refractive index behaviours. Solutions with a molecular concentration 0.28 M or lower remain in a liquid state at room temperature and can be used directly for optical measurements. For high concentration solutions (molar concentration > 0.28 M), the solutions were heated to 65 °C in an oven (OV-480A, BLUE M) to facilitate the dissolution process of the solid and all optical measurements were carried out while the solutions were maintained in a uniform liquid state. It is noteworthy that at high concentrations of dye solutions, the molarity of the dye solutions could significantly be deviated from theoretical concentration assuming the final solution remained at the same concentration with the water added, and a recalibration of molar concentration was generated as shown in **Table 1**. For low concentrations, the volume changes due to added dye powder was minimum thus the prepared molar concentrations were assumed as accurate without further calibration.

### *Preparation of Silica Gel with Dye Molecules*

Scattering phantoms were prepared to demonstrate the capabilities of adjusting scattering properties with dye solutions. The composite samples were prepared by combining low-melting temperature agarose (0.6 mg/mL, 50101, SeaPlaque) with tartrazine powder and the mixture was then dissolved by adding a solution containing 10 mg/mL of 1 μm silica spheres (SISN1000, nanoComposix) to achieve different final dye concentrations following **Table 1**. The preparation method required precise mixing to ensure that all components were well dispersed, preventing formation or uneven spread of silica nanospheres and dye molecules. After combining all of the components, each sample mixture was heated under controlled conditions for 10 minutes at 75°C. After heating, the homogeneity of the solution was checked by eye to ensure complete dissolution of agarose particles, resulting in a clear fluid. Extensive stirring with vortex (2800 rpm, INTLLAB Vortex Mixer) was applied after heating ensuring a consistent distribution of the silica spheres inside the solution.

After heating, 1 mL solution was carefully introduced into a 15×15×5 mm disposable base molds (62352-15, Electron Microscopy Sciences) using volumetric pipettes. This mold gave the gel a well-defined shape, ensuring that they could be compared across concentrations. After the solution was added into the mold, the samples were left to cool at room temperature, allowing the agarose matrix to form the gel without causing heterogeneity of the dye molecules and silica particles within the composite. After gelation, the produced samples were suitable for additional optical characterization.

The images of the scattering phantoms were captured using a compact imaging setup built on tabletop optical table. The system consists of a Thorlabs CMOS camera (CS165CU/M, Zelux) paired with a lens (MVL4WA, Thorlab) featuring 3.5 mm focal length and f/1.4 aperture. A LED light sight source was used for back-illumination, and images were taken using ThorCam software (version 3.7.0.6).

**Table 1**. Concentrations and real refractive index of different tartrazine dye solutions.

| Tartrazine mass (mg) dissolved in 1 mL water | Relative percentage of tartrazine in solution | Calibrated molecular concentration (M) | Peak real refractive index ($n'$) at 490 nm |
|---|---|---|---|
| 53.5 | 5.1% | 0.09 | 1.3664 ± 0.0009 |
| 106.9 | 9.7% | 0.18 | 1.3896 ± 0.0023 |
| 160.3 | 13.8% | 0.28 | 1.4087 ± 0.0010 |
| 213.7 | 17.6% | 0.37 | 1.4268 ± 0.0026 |
| 266.8 | 21.1% | 0.44 | 1.4407 ± 0.0015 |
| 320.8 | 24.3% | 0.53 | 1.4547 ± 0.0027 |
| 374.1 | 27.2% | 0.58 | 1.4724 ± 0.0021 |
| 426.6 | 29.9% | 0.65 | 1.4848 ± 0.0015 |
| 481.7 | 32.5% | 0.72 | 1.5025 ± 0.0034 |
| 535.2 | 34.9% | 0.81 | 1.5163 ± 0.0012 |

***UV-VIS-NIR Spectroscopy and Concentration Recalibration***

A UV-VIS-NIR Spectrophotometer (SolidSpec-3700i, Shimadzu) was used to quantify the optical transparency of the dye solutions and phantoms from deep ultraviolet range to near-infrared range. For transmittance of the dye solutions, we utilized standard quartz cuvette (O.D. 45 mm × 12.5 mm × 12.5 mm, Lab4US) with volume 3.5ml and 10 mm optical path length. After completing baseline correction, 3 mL of dye solution was introduced into the cuvette and positioned it the cell holder on the sample side of the Direct Detector Unit for liquid measurement. The reference side was left empty since all the measurements were taken with respect to air. For scattering phantoms, we utilized the horizontal translational stage. The light path was normal to the sample surface, passing the beam directly through the sample and the transmitted light was then collected by the integrating sphere with detectors positioned on the opposite side.

The absorbance spectra were used to quantify the molar concentration of high concentration dye solutions shown in **Table 1**. The Beer-Lambert law relates the absorption of light to the properties of the material through which light is passing. It is expressed as $A = \epsilon c l$, where $A$ is the absorbance, $\epsilon$ is the molar extinction coefficient, $c$ is the concentration of the solution and $l$ is the path length of light through the solution. Using the Beer-Lambert Law, the molar concentration of each tartrazine solution is calculated by measuring absorbance at specific wavelengths that correspond to its peak absorption. Since the variation of volume can be ignored in the millimolar range, absorbance spectra were first measured for the solution at 0.02 mM, 0.04 mM, 0.06 mM, 0.08 mM, and 0.10 mM to generate a calibration curve. High concentration tartrazine solutions were then diluted to this millimolar concentration range by adding deionized water. Because the spectrometer is sensitive to characterize absorption of dye solutions at low concentrations, the molar concentration of the diluted solutions can be accurately characterized, and the original concentration of the dye molecules can be accurately quantified. This recalibration procedure is essential for extracting correct optical properties of the materials as well as comparing the experimental characterization with the theory.

The imaginary part of refractive index $n''$ for a uniform solution can be calculated using the following relation: $n'' = -\ln(10) Abs / (4\pi d \lambda^{-1})$, where $Abs$ is the wavelength dependent absorbance measured from the UV-VIS-NIR spectrometer, $\lambda$ is the wavelength, and $d$ is the optical thickness of the sample during measurement. For scattering phantoms, the total attenuation coefficient $\mu$ of the sample can also be calculated utilizing $\mu = -\ln(10) Abs/d$. The absorbance of the sample $Abs$ is defined as $Abs = 2 - \log_{10}(T)$, and $T$ is the transmittance of light in unit of percentage.

***Theoretical Modelling of Refractive Index***

The real and imaginary parts of the refractive index are correlated by the Kramers-Kronig relations, which are based on the fundamental principle of causality.[13] The imaginary part of the refractive index characterizes the absorption of light within the material, whereas the real part governs the phase velocity of light propagating through it. The connection between the imaginary and real parts of the refractive index can also be expressed in the wavelength domain,[5]

$$n'(\lambda) - 1 = \frac{2}{\pi}\mathrm{P.V.}\int_0^{\infty} \frac{n''(\lambda')}{\lambda'\left(1-\frac{\lambda'^2}{\lambda^2}\right)} d\lambda'$$

Where $n'$ and $n''$ are the real and imaginary parts of the refractive index, respectively, $\lambda$ is the wavelength at which refractive index is being calculated, and $\lambda'$ is the integration variable. This formulation highlights that the refractive index at specific wavelength depends on the entire spectral range. With our customized MATLAB code, the molar imaginary refractive index $n''$ was calculated from the absorption spectra from UV-VIS-NIR spectrometer at low concentrations, and the refractive index modulation at certain molar concentration can be predicted and compared with direct experimental measurements.

***Ellipsometer Measurements***

An ellipsometer (Alpha 2.0, J.A. Woollam) was used to measure the real ($n'$) and imaginary ($n''$) parts of the refractive indices of dye solutions shown in **Table 1**. Specifically, 4 ml of the solution was added to the disposable base molds (37mm × 24mm × 5mm, 62352-37, Electron Microscopy Sciences). The solution covered the entire bottom of the cavity, creating a flat, reflective air-liquid interface. Reflected light was measured at three angles of 65.01°, 70.01°, 75.00° and the refractive index of the solution was calculated with a single interface semi-infinite reflection model[13] utilizing a customized MATLAB code and averaged between three angles. For all measurements, a 1 s acquisition time was applied for each refractive index spectrum.

**Results**

***Transmittance and Absorption of the Dye Solutions***

The optical transmittance properties of dye solutions are crucial for their various applications. Using tartrazine as an example, this azo dye has a molecular weight of 534.36 g/mol and appears as a distinct yellow powder (**Fig. 1a**).[9] Tartrazine is highly soluble, allowing preparation of solutions with concentrations up to 0.8 M (**Table 1**, **Fig. 1b**). To quantify the optical transmission of different solutions, we employed a UV-VIS-NIR spectrophotometer, which measures optical transmission across a broad wavelength range (**Fig. 1c**).

The characteristic yellow color of tartrazine solutions arises from its absorption peak centered around 428 nm (**Fig. 1d**), making it an ideal colorant for food applications and indicating electronic transitions of the dye molecules in this region. Beyond 500 nm, the dye exhibits minimal absorbance, indicating limited interaction with visible light in this range. By measuring the absorption spectra of low-concentration dye solutions, we established a calibration curve linking absorbance to molar concentration based on Lambert-Beer's law (**Fig. 1e**) and extracted the extinction coefficient of tartrazine as $25.0 \pm 1.3$ $\mathrm{mM^{-1}\ cm^{-1}}$. From these absorption spectra, we also derived the imaginary part of the refractive index for each solution, with **Fig. 1f** showing the molar imaginary part calculated by averaging results from different concentrations in **Fig. 1e**. UV-VIS-NIR spectrometer measurements allow accurate measurement of the absorption peak at low concentrations but also introduces relatively large calculation error when calculating molar absorption, especially in non-absorbing region as shown in **Fig. 1f**. It is important to note that our measurements are constrained by the spectrometer's available wavelengths, excluding accurate measurements of strong absorbance in the lower wavelength range (<300 nm), relating to higher-energy electronic transitions. These gaps in the absorption spectra may introduce errors in the refractive index calculations. At longer wavelengths, water absorption dominates,[14,15] limiting the most valuable range of optical characterizations to 1500 nm.

For high-concentration tartrazine solutions, spectral signatures corresponding to the electronic transitions are not observable due to strong absorption at certain wavelengths. In the transmittance spectra (Fig. 1g), the transmittance of various dye solutions (0.00 M to 0.81 M) sharply decreases and approaches zero at shorter wavelength (< 600 nm), indicating strong absorption in this spectral region. This behavior corresponds to the peak observed in the imaginary part of the refractive index (*n*") at these wavelengths (Fig. 2b). At longer wavelengths (> 600 nm), *n*" approaches zero, resulting in substantially reduced absorption and, therefore, increased transparency of tartrazine solutions. Since dye solutions are primarily composed of more water (~70% or more), the transmittance spectra of tartrazine solutions become similar to those of water for wavelengths where absorption from dye molecules is negligible. This increased transparency at longer wavelengths could be advantageous for applications requiring low absorption, such as biomedical imaging and optical communications.[7,16] Absorbance spectra show that absorbance increases with dye concentration (**Fig. 1h**) but the exact absorbance values are not measurable due to the absorbance limits of the spectrometer. The imaginary part of the refractive index ($n''$) can be calculated from the absorbance values and **Fig. 1i** shows the dependence of $n''$ in a semi-log plot to illustrate the difference in $n''$ as it approaches the near-infrared wavelengths. Beyond 900 nm, $n''$ approaches that of water, demonstrating that tartrazine absorbs little light in the near-infrared range in comparison to water.

***Refractive Index of the Dye Solutions***

Ellipsometry offers a direct method for measuring the refractive indices of dye solutions. In ellipsometry measurements, light is reflected from the sample surface and collected by a detector to analyze the relative change in the polarization of the light (**Fig. 2a**). This technique requires the solution surface to be flat and stable.[17] Using ellipsometry, we measured the optical properties of tartrazine solutions at various concentrations, including the real part ($n'$) and the imaginary part ($n''$) of the refractive index across a wavelength range of 300 nm to 1000 nm. **Fig. 2b** shows measurements for a 0.65 M tartrazine solution, with standard deviations calculated from three independent measurements at different angles.

To compare the experimental measurements with theoretical predictions, we calculated the real refractive index modulation ($\Delta n'$) for both experimental and theoretical results. For the experimental results, we calculated the difference in the real refractive index of the 0.65 M solution compared to water. For the theoretical results, we first estimated the imaginary refractive index of the 0.65 M solution based on **Fig. 1f** and then calculated the refractive index change using the Kramers-Kronig relations. The two results qualitatively match, but the theoretical results show a sharper increase with a lower refractive index change at longer wavelengths (**Fig. 2c**). This discrepancy is likely due to the nonlinear dependence of absorbance on concentration at high concentrations and the incomplete wavelength spectrum in the spectroscopy measurements.

The refractive indices of different solutions were then systematically characterized. **Fig. 2d** displays the imaginary refractive index ($n''$), which is concentration-dependent, with higher concentrations exhibiting greater peak magnitudes. The imaginary part of the refractive index reflects the solutions' absorption behavior, with the peak indicating substantial light absorption in this spectral band. The graphs for various concentrations demonstrate a consistent increase in $n''$ with increasing molarity, confirming that higher concentrations result in greater absorption. However, the dependence of $n''$ on concentration on a fixed wavelength is not exactly linear (**Fig. 2e**) and the peak of $n''$ also shifts to lower wavelength at higher concentration (**Fig. 2d**), suggesting additional effects such as molecular aggregation or fluctuations in light scattering at higher concentrations.

The real refractive index ($n'$, **Fig 2f**) shows a clear peak in the short visible wavelengths (around 400-500 nm) indicating high refractive index values and gradually decays to a plateau as they extend into the long visible range (>700 nm). By varying the concentration of the dye, the refractive index values of the solution can be tuned from approximately 1.332 to 1.521, but the tunability strongly depends on the wavelength of interests. The dependence of $n'$ on concentrations follow a quasi-linear pattern (**Fig. 2g**) and the molar $n'$ change was extracted to be $0.215 \pm 0.004$ $M^{-1}$, $0.153 \pm 0.003$ $M^{-1}$, $0.149 \pm 0.002$ $M^{-1}$ at 500 nm, 600 nm, and 700 nm, respectively.

These ellipsometry measurements highlight the importance of direct optical measurements in quantifying the optical properties of dye solutions. The findings indicate that tartrazine significantly influences the optical characteristics of aqueous solutions, and understanding this effect is important for tuning optical properties of the system. However, a simple linear model measured at low concentrations is insufficient to predict the optical properties of high-concentration solutions (**Fig. 2c, Fig. 2e**). The nonlinear pattern of increase implies that the refractive index does not increase uniformly with concentration. Instead, chemical interactions, aggregation processes, or changes in light absorption may cause deviations from a simple linear dependence. This characteristic is critical for applications in optical sensing and spectroscopy that require precise refractive index control. Further research should investigate how different solvents and pH levels alter tartrazine's optical characteristics.[11] Additionally, studying fluorescence in conjunction with the refractive index would provide a more comprehensive understanding of its spectral properties.

***Light Propagation through Scattering Phantom with Dye Molecules***

To investigate light propagation through a scattering medium, we doped silica spheres into a 0.6% agarose hydrogel and gradually increased the concentration of tartrazine within the scattering medium (**Fig. 3a**). Due to the difference in refractive index between silica particles and hydrogel, the scattering phantom without dye molecules are strongly scattering.[18,19] These composites have been utilized as tissue-mimicking phantoms in biomedical imaging systems at specific optical wavelengths, exhibiting strong scattering due to the higher refractive index of silica particles compared to water.[20] By placing different phantoms on top of a grid or pattern, we observed that the transparency of the composite system utilizing a customized optical system (**Fig. 3b**). The optical transmittance of these phantoms can further be quantified by the UV-VIS-NIR spectrometer while keeping the sample flat on a horizontal stage (**Fig. 3c**).

The light propagation through the scattering medium is non-trivial. We observed that the transparency of the composite system increased with dye concentration from 0.0 M to 0.65 M, allowing visualization of the underlying patterns (**Fig. 3d**). Notably, the transparency of the phantom begins to decrease when the concentration is further raised from 0.65 M to 0.81 M, causing the patterns to blur again. To quantify these differences, a line was drawn at the center of the image across the grid, and the intensity values were monitored along this line (**Fig. 3e**). With increasing dye concentrations, the overall intensity value decreased due to increasing absorption in the composite, while the resolution of the image increased as the sharp features start to emerge. To quantify the resolution differences, a modulation transfer function (MTF) was introduced to quantify the system's resolution by dividing the difference between maximum and minimum intensity values by the sum of maximum intensity values.[21] A high MTF values closer to 1 indicates that features can be clearly resolved, while lower MTF values closer to 0 indicate relatively low resolution. We observed that at the optimized dye concentration, the resolution and MTF value of the image reached a level close to the hydrogel case, where there is almost no scattering. However, the resolution starts to degrade as the dye concentration increases further, suggesting that effective light transmission can only be achieved at certain dye concentration for this scattering medium.

The transmittance of the phantoms was then quantified using a UV-VIS-NIR spectrophotometer in a unique flat geometry. From the transmission results across the broad spectrum, we observed that the maximum transmittance wavelength changes significantly with dye concentration (**Fig. 3f**). By calculating the transmittance enhancement factor, defined as the transmittance ratio between a specific concentration and the original scattering phantom, we found that optical transmission through the scattering medium can increase by over 120 times, with optimized wavelengths also shifting with concentration (**Fig. 3g**). For example, the maximum transmittance value shifts from 600 nm to 800 nm as the concentration changes from 0.65 M to 0.81 M. The exact transmittance values are related to the sample thickness, and to avoid bias due to thickness variations, the normalized attenuation coefficient was calculated (**Fig. 3h**). This coefficient is defined as the ratio between the total attenuation coefficient for phantoms of different dye concentrations and the original scattering phantom without dye, and the attenuation coefficient approaches zero at certain wavelengths indicating the effective light propagation through the scattering medium.

## Discussion

Our results have detailed the optical characterization of dye solutions, providing a comprehensive quantification of the optical properties of the dye solutions for potential applications in bioimaging and optical systems.[7,16] Organic dyes are commonly employed in food and medicines, due to their brilliant pigmentation, solubility, and tunable optical qualities, although concerns about its potential health implications, including as allergic responses and hyperactivity in sensitive people, have prompted additional research into its behavior in various situations.[9] By systematically measuring optical properties including absorption spectra, real and imaginary refractive indices, we have established a robust methodology that highlights the versatility and adaptability of food dye solutions. This characterization is crucial for understanding how these solutions can be optimized for specific optical applications, ranging from enhancing image contrast in biomedical imaging to improving signal clarity in optical communication systems. The detailed optical protocols will also serve as a foundational reference for future research and development in this area.

Based on the principles of refractive index engineering, we can efficiently tune light propagation through scattering media. By adjusting the concentration and combination of food dyes, we can precisely control the refractive index of the solutions, thereby accurately tuning how light interacts with the scattering medium. This tunability allows for the customization of optical environments to achieve desired outcomes, such as minimizing light scattering for light propagation through scattering medium or enhancing light scattering for display technologies.[4,7] The capability of decoupling the material optical properties with other physical properties is critical in a variety of scientific and technical applications. In the growing field of flexible electronics, designing high refractive index soft materials has been challenging and integration of food dyes could be a unique strategy to engineer the optical properties of the materials while maintaining the mechanical softness.[6]

The research also highlights the novel application of food molecules in optical design. Traditionally, optical materials have relied on synthetic compounds that can be expensive and potentially harmful to the environment. In contrast, food dyes offer a sustainable and eco-friendly alternative. Our findings suggest that food molecules can be effectively used to design optical materials with tailored properties, opening up new possibilities for innovation in the field. For instance, the use of food dyes in creating refractive index fluids not only reduces costs but also ensures safety and environmental compatibility. This novel application of food molecules in optical design represents a significant step forward in the development of sustainable optical technologies, with potential implications for a wide range of industries, including healthcare, communications, and environmental monitoring.[22,23]

**Acknowledgements:** ZO acknowledges support from The University of Texas at Dallas.

**Conflict of Interest:** All authors declare no conflict of interests.

**Data and Materials Availability:** All data and materials are available from the corresponding author through reasonable requests.

**Author Contributions:** Conceptualization and supervision: Z.O. Materials characterization and data presentation: M.W.S., S.B., R.K., and S.N. Writing: M.W.S., S.B., R.K., S.N., and Z.O.

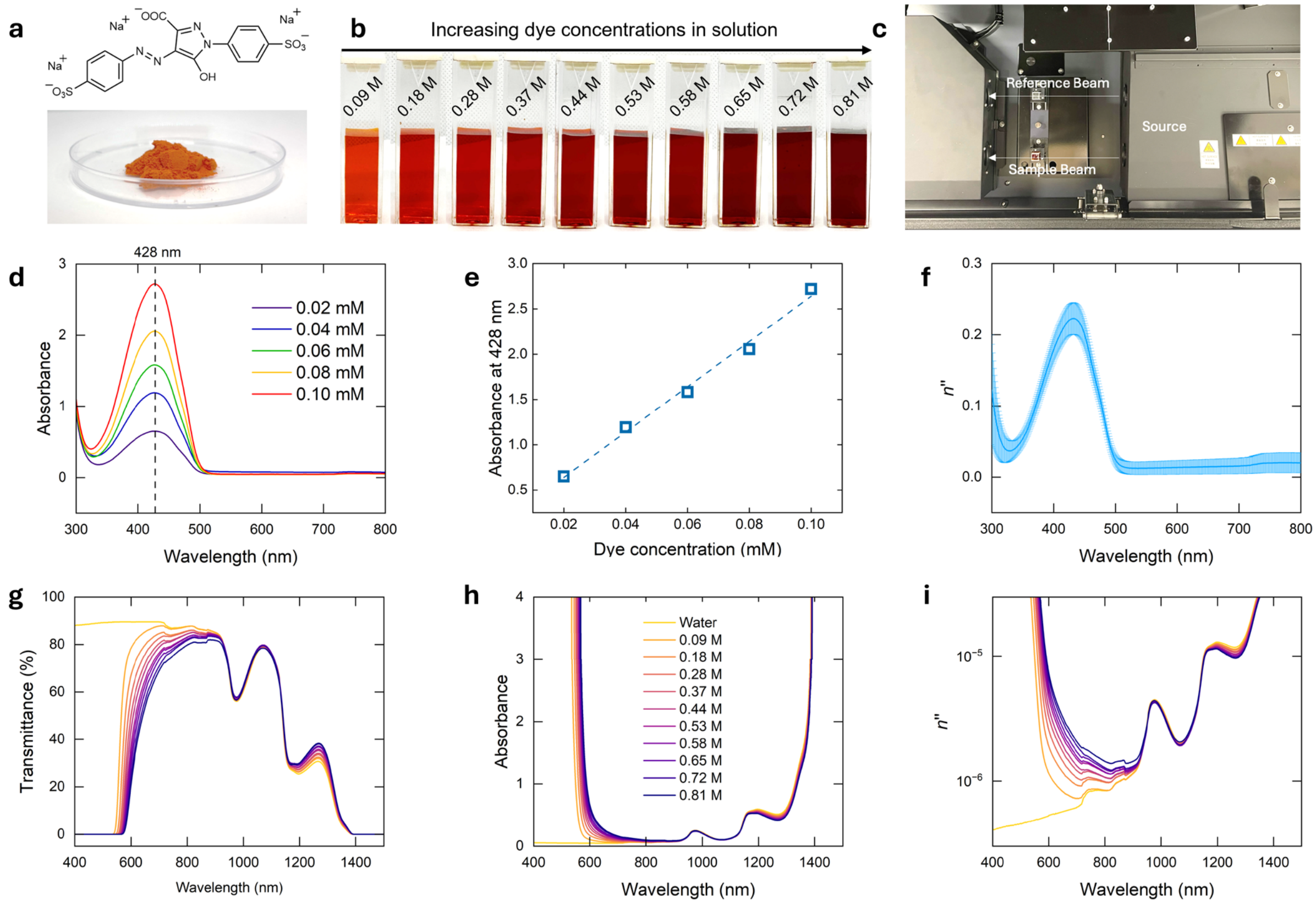

**Figure 1.** Transmittance of dye solutions characterized by UV-VIS-NIR spectrophotometer. (**a**) Chemical structure of the tartrazine and photo of the powder. (**b**) Photo of the dye solutions at different concentrations taken in a 1 cm cuvette. (**c**) Setup to measure the transmittance of the dye solutions using the spectrometer. **(d)** Absorption spectra of dye solutions at low concentration and dashed line indicates the absorption peak at 428 nm. (**e**) Dependence of absorbance and dye concentrations at 428 nm. The dashed line is the linear regression of the experimental data with a slope of 25.0 ± 1.3 $mM^{-1}$ $cm^{-1}$. (**f**) Imaginary part of refractive index ($n''$) for a 1 M solution assuming linear dependence of dye absorption on dye concentration. (**g**) Transmittance, (**h**) absorption, and (**i**) imaginary part of the refractive index ($n''$) for solutions of various high concentration dye solutions. (**g-i**) Different colors of curves share the same notation as shown in (**h**).

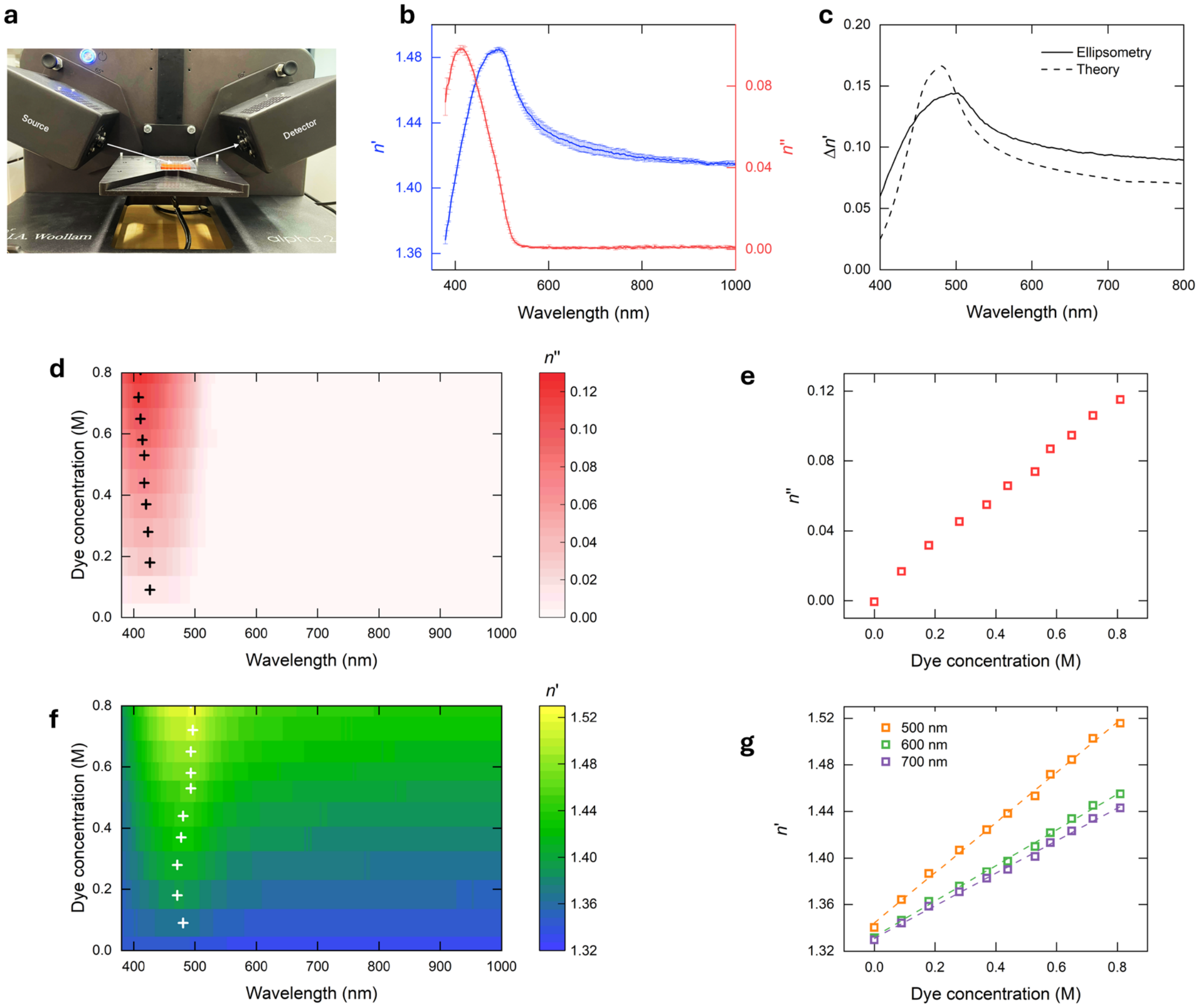

**Figure 2**. Refractive indices of dye solutions characterized by ellipsometer. (**a**) Photograph of the setup to measure the refractive indices of the dye solutions. (**b**) Real (blue) and imaginary (red) part of the refractive index for 0.65 M dye solution. (**c**) Comparison of refractive index modulation between ellipsometry results and theory estimation. (**d**) Imaginary part of the refractive index ($n''$) for dye solutions at various concentrations and peak locations for each concentration are marked by the black crosses. (**e**) Dependence of $n''$ on dye concentration at fixed wavelength of 427 nm. (**f**) Real part of the refractive index ($n'$) for dye solutions at various concentrations and peak locations for each concentration are marked by the white crosses. (**g**) Dependence of $n'$ on dye concentration at fixed wavelength of 500 nm, 600 nm, and 800 nm. The dashed lines are linear regression of the data points, from which the molar refractive index change coefficients are extracted.

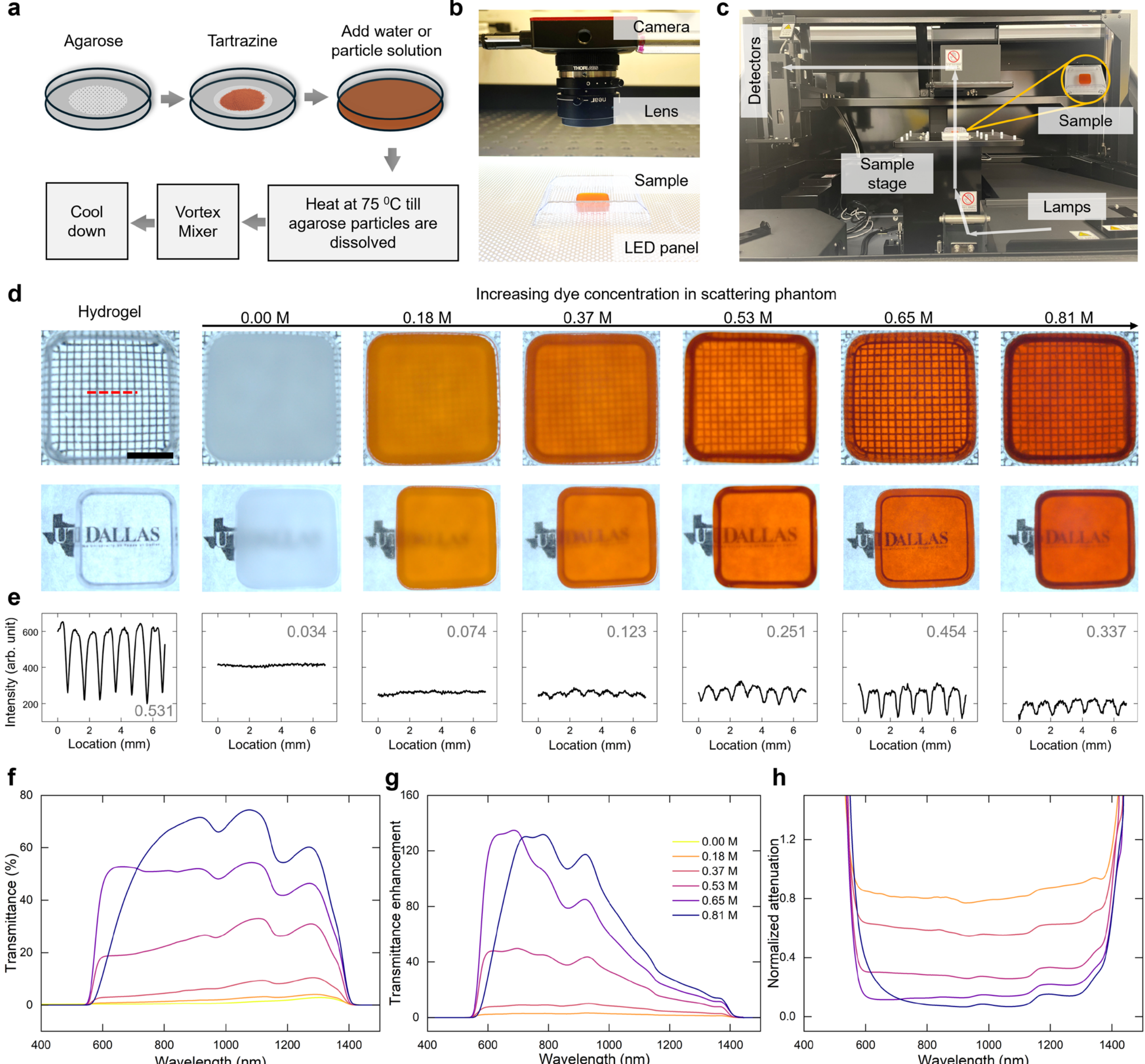

**Figure 3**. Controlling light propagation through scattering phantoms by changing dye concentrations. (**a**) Schematic illustration the preparation of scattering phantoms containing dye molecules. (**b**) Photograms of the optical set up to visualize the pattern through a phantom. (**c**) Optical path illustrating the measurement of light transmittance through phantoms. (**d**) Photographs of different phantoms placed on top of 1 mm grid and pattern. The red dashed line indicates the location of line scan for intensity measurements. Scale bar: 5 mm. (**e**) Line intensity profiles across different grid images and the values on the plot are the modular transfer function (MTF) calculated based on intensity variations. (**f**) Transmittance, (**g**) transmittance enhancement, and (**h**) normalized attenuation at different wavelengths for different phantoms. (**f-h**) Different colors of curves share the same notation as shown in (**g**).

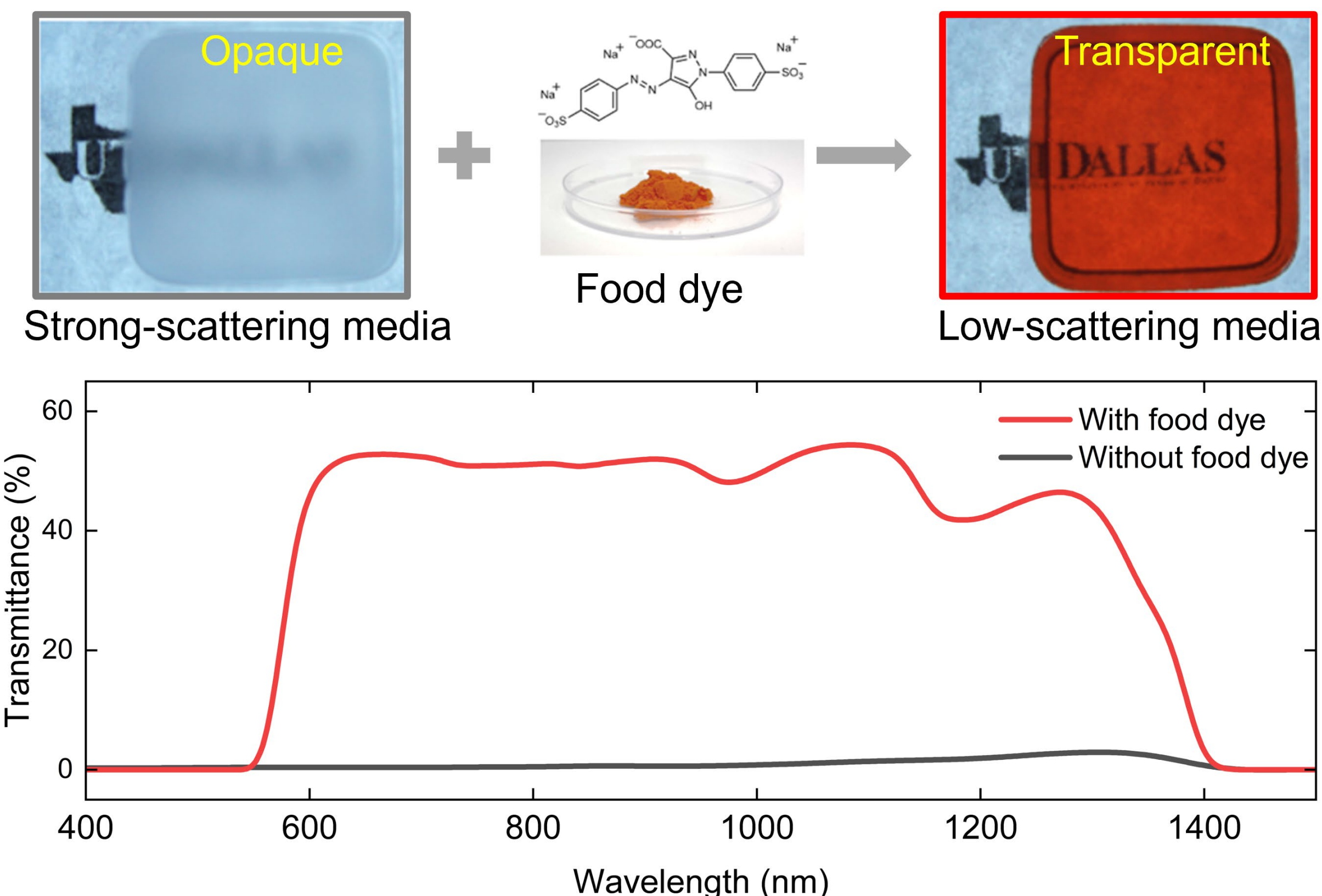

**Graphical abstract**